\documentclass[12pt,aps,amssymb,amsmath,showpacs,prb]{revtex4}
\usepackage{graphicx}
\usepackage{dcolumn}
\usepackage{bm}
\usepackage{natbib}

\begin{document}

\title{Spectral-weight function of the ionic Hubbard model}
\author{M.C. Refolio}
\email{refolio@imaff.cfmac.csic.es}
\author{ J.M. L\`{o}pez Sancho}
\author{J. Rubio}

\affiliation{Instituto de Matem\`{a}ticas y F\`{\i}sica
Fundamental, CSIC\\
Serrano 113 bis, E28006 Madrid, Spain}
\date{\today}

\begin{abstract}
The one-electron spectral-weight function of the half-filled ionic Hubbard
model is calculated by means of Quantum Monte Carlo. A metallic regime
occurs between two values of the coupling constant (Hubbard $U$) $%
U_{1}<U_{2} $. The system is a band insulator below $U_{1}$, with a strong
charge density wave corresponding to a large ionicity, and an increasingly
antiferromagnetic (AF) insulator above $U_{2}$ evolving into a Mott
insulator as $U\rightarrow \infty $ . The intermediate regime, which is both
AF and dimerized, \ is caused by a two-peak structure at $ka$=$0.5{\pi }$
(the Fermi surface). As $U $ increases both peaks approach each other,
overlap, and separate again, the system becoming metallic in the overlap
region. This behavior can be traced to a site-parity change of the ground
state(GS), corresponding to a curve-crossing between two GS of different
site parity.
\end{abstract}


\maketitle

\section{ Introduction}

In quasi-one dimensional materials like halogen-bridged transition metal
chain complexes, conjugated polymers, organic charge transfer salts, or
inorganic blue bronzes, the itinerancy of the electrons strongly competes
with electron-electron and electron-phonon interactions which tend to
localize the charge carriers by establishing commensurate spin or
charge-density wave ground states (GS). At half filling band (BI) or Mott
(MI) insulating phases are favored over the metallic state. Quantum phase
transitions between the insulating phases are possible and the character of
the electron excitation spectra reflect the properties of the different GS.
A controversial issue is the nature of the BI-MI transition as well as
whether or not only one critical point separates both insulating phases in
purely electronic model Hamiltonians\cite{resta, fabrizio}. Despite their
importance in low dimensional materials\cite{jeckel,fehske} phonon dynamical
effects will be ignored in this paper which will be concerned with the
so-called one dimensional (1D) ionic Hubbard model (IHM), just a Hubbard
model with a staggered potential which displaces by $\Delta $ the energy
levels of even and odd sites. It is described by the Hamiltonian

\begin{equation}
H=-t\sum_{<ij>s}c_{is}^{+}c_{js}+\frac{\Delta }{2}\sum_{is}\left( -1\right)
^{i}n_{is}+U\sum_{i}\left( n_{i\uparrow }-\frac{1}{2}\right) \left(
n_{i\downarrow }-\frac{1}{2}\right) ,
\end{equation}
where electrons of spin $s=\pm \frac{1}{2}$ hop with amplitude t between
nearest-neighbor inequivalent sites of energy level $\pm \frac{1}{2}\ \Delta
$ ( $\Delta =1$ in what follows). Electrons of opposite spin on the same
site feel a repulsion $U$.

This model Hamiltonian, originally proposed by Nagaosa \cite{naga} and later
by Egami\cite{ega} as a model ferroelectric, is ideal for studying the issue
of quantum phase transitions in electron systems. On general grounds one
expects a transition to occur from an ionic band insulator to a
strongly-correlated Mott insulator\ as $U$ increases. Evidence for such a
transition was found in exact-diagonalization calculations by Resta et al
\cite{resta} predicting a metallic point at a critical value $U=U_{c}$
separating both insulating phases. This transition is signaled by a
polarization jump with a sign change associated with a site parity change of
the GS. The same conclusion was supported by Gidopoulos et al\cite{gido},
who showed the reversal of site parity above to be of magnetic origin, thus
leading to a vanishing of the spin gap, $\Delta _{s}=0$. Likewise, Brune et
al\cite{brune} by means of the bosonization technique along with Lanczos and
density matrix renormalization group (DMRG) calculations lent support to
this picture.

On the other hand, Fabrizio et al\cite{fabrizio} shed new interest into the
IHM in a field-theoretical bosonization analysis where they propose a new
scenario for the intermediate region, $U\simeq \Delta $, between the BI and
the MI. The model should instead exhibit two quantum phase transitions: one
from a BI state to a long-range bond-ordered (BO) state predicted to be in
the Ising universality class and a second one from the BO to the MI state
predicted to be a Kosterlitz-Thouless transition. Such transitions to BO
states have recently been found in 1D Hubbard models with extended
interactions (U-V) by Nakamura\cite{naka}. The BO state is a broken symmetry
state in which the system becomes ferroelectric due strictly to
electron-electron interactions even if all the atoms are at centers of
inversion. This is also called the spontaneously dimerized insulating (SDI)
phase, which, as argued by Fabrizio et al\cite{fabrizio}, implies a finite
spin gap, $\Delta _{s}$ $\succ 0$. Such a SDI phase has been observed
numerically by Wilkens and Martin\cite{wilkens} in a variational quantum
Monte Carlo study. More precisely, they found one transition from the BI to
the correlated SDI phase, but no second transition to the MI phase which is
approached asymptotically by the SDI phase as $U\rightarrow \infty $. Recent
DMRG calculations \cite{yasu,qin} lead to mutually conflicting results. Thus
whereas Takada and Kido\cite{yasu} support the Resta et al\cite{resta}
picture of a single first-order transition, Qin et al\cite{qin} support the
Fabrizio et al \cite{fabrizio} scenario of two second-order quantum phase
transitions.

In this paper we present the one-electron spectral-weight function A(k$%
\omega $) of the IHM. The calculation is done by means of a Quantum Monte
Carlo (QMC) simulation in the grand canonical ensemble suplemented\ by an
approximate effective action. Its eigenvalues and eigenvectors alow us to
construct explicitly dynamic correlation functions in real time or
frequency. This procedure leads to spectral functions with a reasonably high
resolution. The calculation is performed at a relatively large inverse
temperature, $\beta t=10$ ($t=1$ in what follows), which brings the system
very close to its GS properties. Our calculated A(k$\omega $)clearly shows
metallic character (non-vanishing density of states in the gap) at $%
ka=0.5\pi $ ($a=1$), between two critical U values instead of a single
metallic point at the first threshold. Thus a double BI-M-MI smooth
transition is predicted, the intermediate metallic regime at U$\simeq
2\Delta $ being both antiferromagnetic (AF) and dimerized

\section{ Effective Hamiltonian and correlation functions}

In our calculations, we use the QMC approach in the grand canonical
ensemble, which has been explained in detail by Hirsch\cite{hirsch,hirsch2}
and White et al\cite{white}. As is well-known, the partition function is
factorized into $L$ time slices of extent $\Delta \tau =\beta /L$. For
Hubbard models, the interaction $U\sum_{i}n_{i\uparrow }n_{i\downarrow }$ is
transformed into a discrete Ising field\cite{hirsch}, $\sigma _{il}=\pm 1$,
which depends on lattice site $i$ and imaginary time $\tau =l\Delta \tau $ $%
(l=1,.....L)$. The non-interacting Hamiltonian for a given Ising
configuration (a definite allocation of $+1$ and $-1$ values for all the $%
\sigma _{il}$ is) reads\cite{hirsch2}
\begin{equation}
H(\tau )=\sum_{<ij>s}c_{is}^{+}h_{ij}^{s}\left\{ \sigma _{i}\left( \tau
\right) \right\} c_{js}\equiv H_{l}
\end{equation}
where
\begin{equation}
h_{ij}^{s}\left\{ \sigma _{i}\left( \tau \right) \right\} =t_{ij}+\left(
\Delta \left( -1\right) ^{i}+\frac{U}{2}-\mu +\lambda \alpha \sigma
_{il}\right) \delta _{ij}
\end{equation}
The first term on the rhs of Eq(3) is just the hopping term of the IHM,
Eq(1). In the second term, $\mu $ is the chemical potential, $\alpha =\pm 1$
for up/down electron spin, and $\lambda $ is a non- linear function of $U$
given by $\cosh \lambda \Delta \tau =\exp \left( U\Delta \tau /2\right) $.In
the limit $\Delta \tau \rightarrow 0$ the partition function can be written
as
\begin{equation}
Z=\sum_{<\sigma >}Tr\text{ }\Pi _{l}\text{ }e^{-\Delta \tau
H_{l}}=\sum_{<\sigma >}Tr\text{ }e^{-\Delta \tau
\sum_{l}H_{l}}=\sum_{<\sigma >}Tr\text{ }e^{-\beta (K+V)}
\end{equation}
where $\sum_{<\sigma >}$ means summing over the Ising configurations and $Tr$%
denotes taking the trace over the electron degrees of freedom. $K$ is the $%
\tau $-independent part of $H(\tau )$ and
\begin{equation}
V=\frac{\lambda }{L}\sum_{il}\sigma _{il}\left( n_{i\uparrow
}-n_{i\downarrow }\right) =\lambda \sum_{i}<\sigma _{i}>\left( n_{i\uparrow
}-n_{i\downarrow }\right)
\end{equation}
$<\sigma _{i}>=\left( 1/L\right) \sum_{i}\sigma _{il}$ being the average
Ising spin at the $i$-th site. Therefore, $H_{eff}=K+V$ is $\tau $%
-independent in the limit $\Delta \tau \rightarrow 0$. We adopt this
effective $\tau $-independent Hamiltonian, for doing measurements even if
the number of time slices used in the calculation is finite. To minimize as
much as possible size effects due to a finite $L$, we shall increase $L$ and
seek for convergence. Needless to say, the update of the Ising field is done
following the standard procedure of, say, Ref 15.

The corresponding Hamiltonian matrix for each spin $s$ (Eq(3) with $<\sigma
> $instead of $<\sigma _{il}>$) is now diagonalized and, in terms of its
eigenvalues $\varepsilon _{\mu s}$ and eigenvectors $\mid \mu s>$, we can
construct correlation functions in real time or frecuency. For instance
\begin{equation}
G_{ij}^{s}(\omega )=\sum_{\mu }<i\mid \mu s>\left\{ \frac{1-f_{\mu s}}{%
\omega -\varepsilon _{\mu s}+i\eta }+\frac{f_{\mu s}}{\omega -\varepsilon
_{\mu s}-i\eta }\right\} <\mu s\mid j>=G_{ij}^{s>}(\omega
)+G_{ij}^{s<}(\omega )
\end{equation}
for the causal Green's functions and likewise for other quantities. As
customary, $f_{\mu s}=\left( e^{\beta \varepsilon _{\mu s}}+1\right) ^{-1}$%
is the Fermi-Dirac distribution and $\eta$ a vanishingly small
imaginary part. All these quantities must now be averaged over the
accepted Ising configurations to obtain the final result. It is
just in this averaging process where the correlation effects are
restored. Since the final result is translationally invariant, one
can Fourier transform and find the spectral-weight function (SWF)
\begin{equation}
A_{ks}(\omega )=-\frac{1}{\pi}Im\left\{G_{ks}^{>}(\omega
)-G_{ks}^{<}(\omega )\right\},
\end{equation}
and the DOS
\begin{equation}
N_{s}(\omega )=\frac{1}{N}\sum_{k^{\prime }}A_{ks}(\omega)
\end{equation}
which can also be obtained from (6) as $(1/N)\sum_{i}N_{i}(\omega )$, $%
N_{i}(\omega )$ being the corresponding imaginary part of $G_{ii}^{s}(\omega
)$. For our case of the IHM, this last route allows to calculate the partial
DOS for even/odd atoms by simply restricting the sum over even/odd sites. As
noticed earlier, this building procedure leads to a good resolution of
spectral features.

\section{Double insulator-metal-insulator transition}

We characterize the system by its one-electron SWF, $A_{k}(\omega )$, along
with the charge and spin structure factors, $c_{k}$ and $s_{k}$, defined as
the Fourier transform of the static charge and spin, respectively,
correlation functions. These are given as usual by $c_{ij}=<q_{i}q_{j}>$ and
$s_{ij}=<s_{iz}s_{jz}>$, $q_{i}=n_{i\uparrow }+n_{i\downarrow }$ and $%
s_{iz}=(1/2)(n_{i\uparrow }-n_{i\downarrow })$ being the charge and spin at
the $i-th$ site and $\tau =0$. This type of characterization is useful and
can be of help in understanding the electronic properties of a system.\ For
instance, the SWF tells us the presence or absence of gaps or pseudogaps as $%
U$ increases, which is of fundamental importance in establishing the
insulating or metallic character and possible transitions between them of,
say, the IHM. Likewise, the evolution of $c_{k}$ and $s_{k}$ with $U$,
especially theirs peak at $k=\pi $, gives information about the charge or
magnetic origin of the transitions.

\subsection{The spectral-weight function.}

\begin{figure}
\begin{center}
\includegraphics{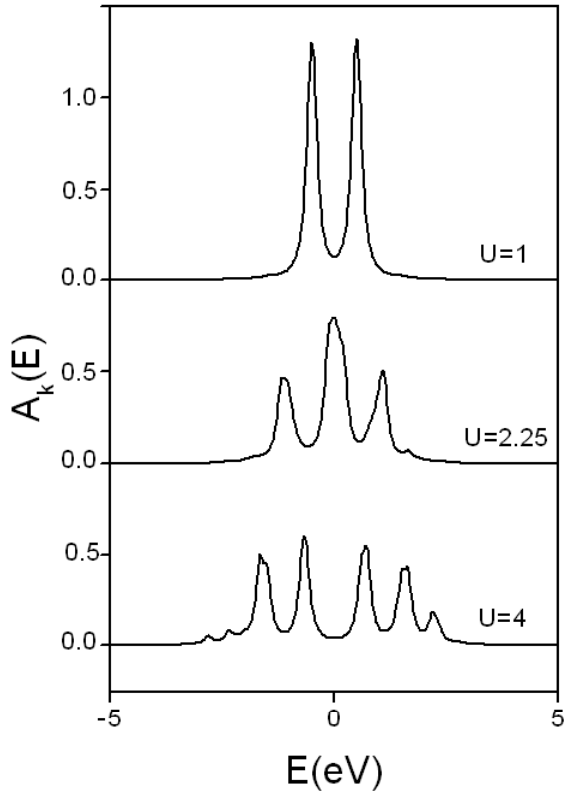}
\caption{The spectral-weight function $A_{k}(\omega)$ for $k=\pi
/2$ and U/t (t=1) values lying in the three coupling regimes of
the ionic Hubbard model. The QMC calculation has been made for a
periodic chain of eighty sites an $\beta=10$}
\end{center}
\end{figure}

As hinted before, it is just the structure of $A_{k}(\omega )$ at $k=\pi /2$
which accounts for the insulating or metallic character of the IHM. Hence,
Fig 1 shows this quantity for three values of $U$ (recall that $t=1$) so
selected to exhibit the three fundamental regimes of the model. The
calculation has been done for a chain of eighty sites with periodic boundary
conditions at half filling and $\beta =10$. Typically 150 time slices, two
hundred warm sweeps and one thousand measurements were taken. At $U=1$, a
two-peak structure with a clear gap in between signals an insulating system.
At $U=2.25$, both peaks have overlapped developing a third peak just at the
midpoint of the gap (which becomes the Fermi level). This three-peak
structure clearly shows a metallic system. Finally, at $U=4$ the central
peak splits into two peaks leaving a gap in between. This four-peak
structure signals again an insulating system which must be somehow of a
different kind from the initial one (see below). As $U$ increases (not
shown) the central gap widens and the four peaks go over into a two-peak
structure as $U\longrightarrow \infty $, the system becoming then a Mott
insulator.

\begin{figure}
\begin{center}
\includegraphics{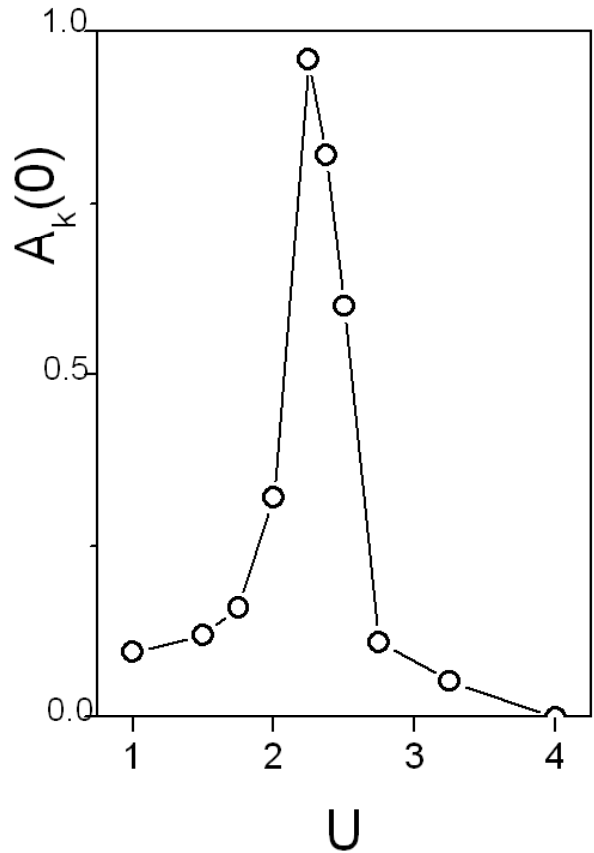}
\caption{$A_{k}(\omega)$ at $k=\pi /2$ (a=1) and E=0 (midgap which
becomes the Fermi level) as U increases. Same calculations of Fig.
1 }
\end{center}
\end{figure}

In order to estimate the approximate extension of the metallic region, Fig 2
shows the evolution of the spectral density at $k=\pi /2$ and $\omega =0$
(the midgap) as $U$ increases. We see a maximum at $U_{m}=2.25$ an onset
threshold at $U_{1}\simeq 1.75$ and a falloff with an approximate endpoint
at $U_{2}=$ $2.75$ eV leaving a zone of metallicity $U_{2}-U_{1}=1$. Since
this quantity may be size-dependent, we proceed to a finite-size scaling in
order to approach the thermodynamic limit. Calculations made on chains of
8,12,16,80,120, and 160 sites show that $U_{2}-U_{1}=1$ is roughly
independent of size, but $U_{m}$ tends to go up, extrapolating to 2.50 when $%
1/N\longrightarrow 0$. Notice that only finite chains of $N=4m$ sites should
be used since only them have $k=\pi /2$ as an allowed wave vector. Chains of
other lengths fail to detect the metallicity. In the thermodynamic limit, of
course, all chains have a wawe vector infinitessimally close to $\pi /2$

\begin{figure}
\begin{center}
\includegraphics{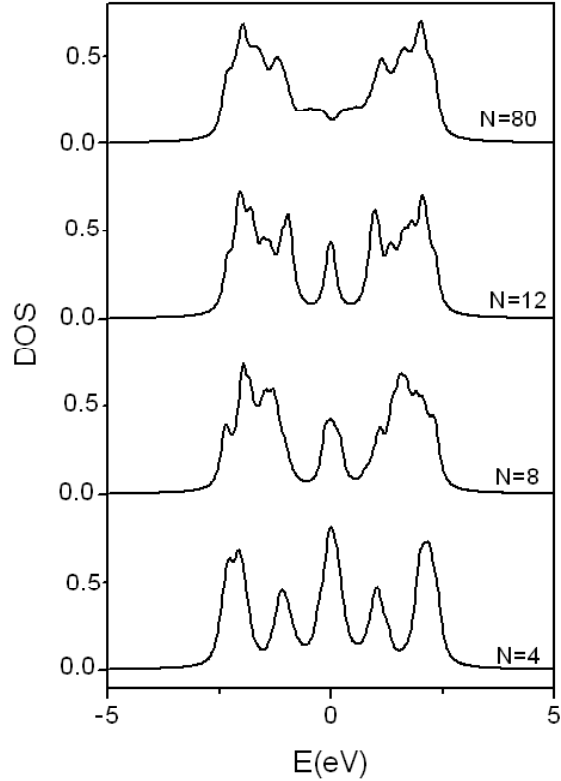}

\caption{Density of states for the same chain as Fig 1 for U and
number of sites N=8,12,16 and 80.}
\end{center}
\end{figure}
The three-peak structure responsible of the metallic phase
disappears as soon as one departs from $k=\pi /2$, showing a
single mainly occupied peak
at $k=(\pi /2)-\delta $ and a mainly empty peak at $k=(\pi /2)+\delta $ ( $%
\delta =2\pi /N$, $N$ being the number of sites). Hence the metallic peak
should be visible in angle-resolved photoelectron spectroscopy and its
inverse, but not in integrated measurements which give the DOS, Eq (8).
Since the weight of that structure falls with $N$ (Fig 3), only a pseudogap
with a sizeable DOS at the Fermi levels should be seen.

\subsection{The charge and spin structure factors}

\begin{figure}
\begin{center}
\includegraphics{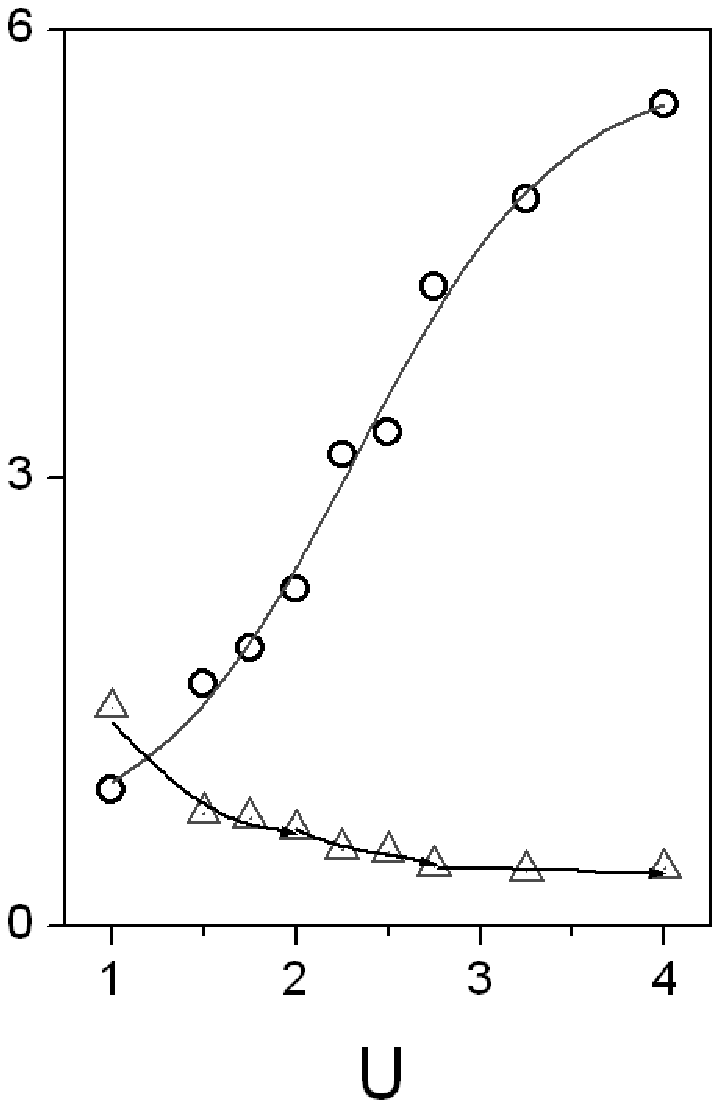}
\caption{The peak at $k=\pi$ of the charge, $c_{k}$ (triangles),
and the spin, $s_{k}$ (circles), structure factor, as U increases,
for the 80-sites periodic chain.}
\end{center}
\end{figure}

Fig 4 finally displays the evolution of the peak of $c_{k}$ and
$s_{k}$ at $k=\pi$ for increasing U. The charge factor
consistently drops from a high value at $U=1$, which signals a
strong CDW in correspondence with a large ionicity (the fractional
ionic character of the bond is $FIC=0.32$) which also falls down
monotonically. The system, therefore starts as an ionic band
insulator at small U. The spin structure factor starts close to
zero, corresponding to a paramagnetic insulator, and monotonically
increases with U, signaling an AF SDW of increasing amplitude. For
large U, the systems evolves into a Mott insulator with a strong
AF SDW and a vanishingly small CDW, i.e., zero ionicity (neutral).

In the intermediate regime, $U_{1}<U_{2}$, the system is both increasingly
AF and neutral (decreasing ionicity, i.e., decreasing CDW). On the other
hand, Table I displays a small sector of the density matrix ($%
n_{ij}^{s}=<c_{is}^{+}c_{js}>$), for $U=2.25$ just the top and left $4\times
4$ matrix, which is sufficient to show that the metallic state of this
intermediate regime is dimerized. Thus $n_{12}^{s}>n_{23}^{s}$, etc,
indicating that the sites ($12$), ($34$), and so on, form dimers. These
dimers weaken for $U>U_{2}$ and tends to disappear as $U\rightarrow \infty $

\section{CONCLUDING REMARKS}

Our QMC study shows three coupling regimes in the ionic Hubbard model: ($i$)
An ionic band insulator below $U=U_{1}$.($ii$) An increasingly
antiferromagnetic and dimerized metal which is gradually becoming neutral in
the intermediate coupling regime $U_{1}<U<U_{2}$. And ($iii$) an
increasingly AF and neutral insulator whose dimerization, measured by $%
n_{12}-n_{23}$, is gradually dropping away for $U>U_{2}$. This last regime
tends to a Mott insulator a $U\rightarrow \infty $.

A remark is here in order. Although the spectral-weight function at $k=\pi /2
$ seems to hint a critical separation among the three coupling regimes (a
threshold and an endpoint), the smooth behavior of all the integrated
quantities (DOS, ionicity, energy, charge and spin structure factors, etc.)
rather suggest a smooth interpolation between the three regimes which should
be distinguished from true quantum phases with sharp, critical, separation
among them.

To conclude, let us finally say that the results of our QMC
calculations would be quite similar to those of Resta et al
\cite{resta}, Wilkens and Martin\cite{wilkens},etc, if the
metallic zone reduced to a single point. On the other hand, they
would also be quite close to those of Fabrizio et al
\cite{fabrizio} should the intermediate regime reported in this
letter turn out to be insulating.

\bigskip

We acknowledge the financial support of the Spanish DGICYT through Project N%
PB 98-0683

\bibliographystyle{unsrt}
%\bibliography{ejemplo}

\bigskip \newpage

\bigskip \bigskip TABLE I. \ \ \ Upper-left $4\times 4$ sector of the
density matrix $n_{ij}^{s}=<c_{is}^{+}c_{js}>$

\bigskip

\smallskip

\bigskip\ \ \ \ \ \ \ \ \ \ \ \ \
\begin{tabular}{cccc}
\hline
\multicolumn{1}{r}{0.5477} & \multicolumn{1}{r}{\ \ 0.2928} &
\multicolumn{1}{r}{\ \ 0.0109} & \multicolumn{1}{r}{\ -0.0555} \\
\multicolumn{1}{r}{0.2928} & \multicolumn{1}{r}{0.4523} & \multicolumn{1}{r}{
0.2116} & \multicolumn{1}{r}{-0.0133} \\
\multicolumn{1}{r}{0.0109} & \multicolumn{1}{r}{0.2116} & \multicolumn{1}{r}{
0.5477} & \multicolumn{1}{r}{0.2928} \\
\multicolumn{1}{r}{0.0555} & \multicolumn{1}{r}{-0.0133} &
\multicolumn{1}{r}{0.2928} & \multicolumn{1}{r}{0.4523} \\ \hline\hline
\end{tabular}

\bigskip

\begin{center}
\medskip \bigskip

\smallskip

\smallskip
\end{center}

\end{document}